# An Inexpensive Arterial Pressure Wave Sensor and its application in different physiological condition

**Shantanu Sur[1] and S. K. Ghatak[2]**
[1] School of Medical Sc & Technology,
[2] Dept of Physics and Meteorology,
Indian Institute of Technology, Kharagpur-721302, INDIA

**Abstract**

Arterial Blood Pressure wave monitoring is considered to be important in assessment of cardiovascular system. We developed a novel pulse wave detection system using low frequency specific piezoelectric material as pressure wave sensor. The transducer detects the periodic change in the arterial wall diameter produced by pressure wave and the amplified signal after integration represents the pressure wave. The signal before integration is proportional to the rate of change of pressure wave and it not only reproduces the pressure waveform faithfully, but also its sharper nature helps to reliably detect the heart period variability (HPV). We have studied the position-specific (e.g. over carotid or radial artery) nature of change of this pulse wave signal (shape and amplitude) and also the changes at different physiological states.

# 1.Introduction:

Arterial pulse examination has long been considered to be important in assessment of cardiovascular system. The dynamical measurement of pulse waveform drew considerable attention for past few decades because of the relationship of specific change of waveform contour with pathological conditions. Numerous methods based on various physical principles were developed to faithfully reproduce the arterial blood pressure waveform. Direct blood pressure monitoring with an arterial catheter is presently considered to be the most accurate method but being an invasive procedure it has various disadvantages including patient discomfort, demand of skilled professional (for catheter insertion) and chance of complication. Oscillometric method is a common method to assess Blood Pressure automatically in a noninvasive way, but does not provide the wave shape.

Currently research is more focused on noninvasive determination of pressure waveform from peripheral artery (radial artery) and synthesis of aortic characteristics from it [1-3]. The periodic contraction of heart and distensibility of the arterial wall results in the pressure wave which travels from aorta to periphery. One commonly used principle to obtain the pressure waveform is applanation tonometry using a strain gauze sensor placed over a pulsating artery. Other method utilizes volume clamp principle of Jan Peñáz and an infrared photoplethysmographic finger cuff is used to measure the pressure waveform [4,5]. Although this method allows beat to beat measurement of blood pressure, due to distortion of waveform assessment from peripheral site such as finger shows wide variability particularly in critical situations such as induction of anesthesia [6,7].

Heart Period Variability (HPV) is conventionally measured from ECG. However there are few recent studies where HPV measured from the systolic blood pressure (SBP) peak intervals are shown to closely resemble that of ECG RR interval [8,9]. One potential error is the slow rise and decline of the systolic blood pressure wave, which is produced mechanically due to interaction of blood and arterial wall, thus may be difficult to locate the exact SBP peak. However the pulse wave signal we get using our sensor represents the rate of change of pressure development in the artery thus have a much sharper peak and promises to serve as better measure to HPV. Another source of potential error could be due to Pulse Wave Velocity (PWV) which represents the speed by which a pressure pulse moved from central aorta to periphery. PWV is highly dependent on structural and functional property of arterial tree, as for example it increases in conditions like arteriosclerosis. But it is independent of heart diseases and its velocity is not affected by fixed or altered heart rate [10] and thus unlikely to induce error in measuring HPV.

Here we have developed a transducer system and carried out experiments to assess the sensitivity and accuracy of the Piezoelectric pressure wave sensor in detecting the changes at different physiological conditions and whether the measured pulse wave signal (representing the rate of change of pressure wave) can provide equivalent information as obtained from conventional method of pressure wave measurement.

## 2. Sensor Assembly and Measurement of Pressure Pulse

The block diagram of sensor system is sketched in Fig.1a. The transducer is piezoelectric crystal encapsulated with surface area ~0.8 $cm^2$. The electrical signal generated by transducer is passed through notch filter that eliminates power line frequency (50Hz) signal and then amplified, displayed and stored in computer. Entire assembly with transducer positioned over artery of subjects is shown in Fig.1b. The signal is integrated by electronic integrator and can be displayed and stored in another channel. The signal produced when transducer responds to pulsation of artery is proportional to time derivative of pressure wave and integrated signal is thus measure of pulse pressure.

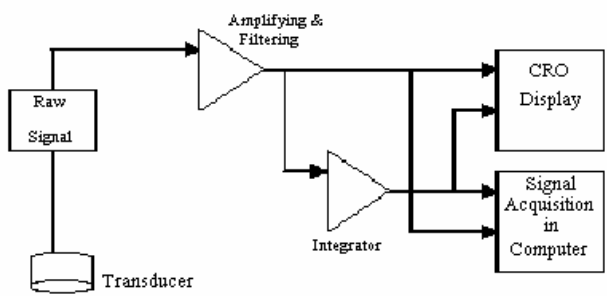
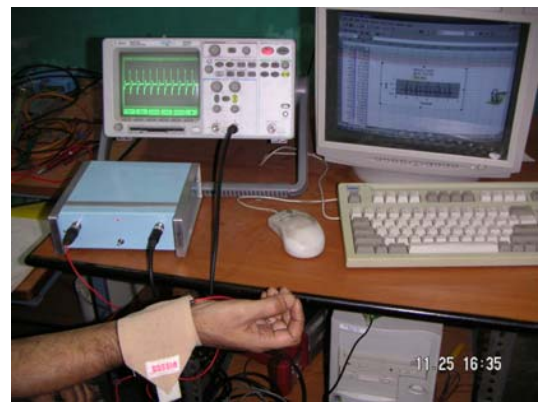

Fig. 1a                           Fig. 1b

Fig.1: a) block diagram of sensor  b) Complete assembly for pressure-pulse measurement with sensor positioned on artery of subject

To record the blood pressure waveform the transducer was placed gently over the artery after palpating it by fingers. For long, continuous and stable recording the sensor was further fixed to limb by the use of an elastic strap (e. g. while recording from Radial artery, Brachial artery). To get the signal from the internal carotid artery the sensor was placed gently over the artery by hand. In all cases data obtained in first few minutes were discarded to avoid the possible instability of patient's pressure and instrumental drift. During recording, care was taken to restrict movement of the limb or part from where the recording was done in order to reduce unwanted noise. The study group consisted of healthy volunteers of age group between 25-60 years.

**3.Results**

The piezo-electric transducer records the rate of change of blood pressure (dp/dt) in the artery of interest. The true pressure waveform can be obtained by passing the signal through an integrator (Fig.-2). The original signal contains a distinct positive and a negative phase with a characteristic separation of the phases depending on the position of recording while the integrated signal have a pressure peak and a dicrotic notch which corresponds to aortic valve closure.

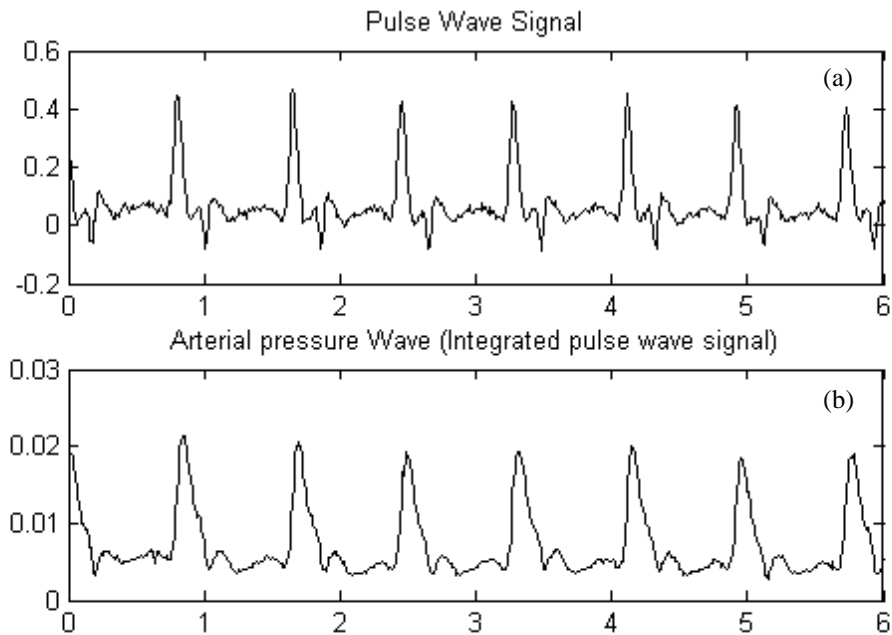

Fig.- 2: (A) Shows the signal obtained from carotid artery by the sensor used. There are one positive and one negative peak and a flat portion between the peaks. The positive peak denotes the increase in the blood pressure and the negative peak denotes the reduction in a cardiac cycle. (B) The same signal after integration represents arterial blood pressure waveform.

The pressure wave signal was used to analyze the heart rate variability. For this purpose 100 seconds recording was done (sampling rate 200 Hz) and the time bins of positive peaks were determined. Mean Heart Rate, STD, RMSSD (Root Mean Squared

Successive Differences), NN50 (instantaneous difference over 50 ms between two consecutive normal-to-normal RR interval) were determined using HRV analysis software ( A free software developed by The Biomedical Signal Analysis Group, Department of Applied Physics, University of Kuopio, Finland). Simultaneous digital ECG recordings were done and extracting the R wave peaks in the ECG performed similar statistical analyses. The ECG and pressure wave data shows very close corroboration (n=3). (Table1) Frequency domain analysis (Autoregressive analysis) of RR interval and pulse wave peak-peak interval was also done by the same HRV analysis software. Analysis shows similar positioning of Low Frequency (0.04-0.15 Hz,LF) and High Frequency (0.15-0.40 Hz,HF) peaks although the power differed considerably.(Fig 3)

|  | Pulse Wave signal peak-peak data | ECG R-R Data |
|---|---|---|
| Heart Rate | 73.24 +/- 7.48 | 73.14 +/- 7.19 |
| STD | 3.45 +/- 1.96 | 3.86 +/- 2.36 |
| RMSSD | 33.43 +/- 14.79 | 33.30 +/- 18.46 |
| NN50 | 14.00 +/-11.04 | 10.67 +/- 9.46 |
| pNN50 | 13.37 +/- 10.07 | 10.13 +/- 8.54 |

Table 1: Table shows Time Domain analysis of Heart Rate Variability detected by Pulse Wave Signal Peak-Peak data and ECG R-R data.

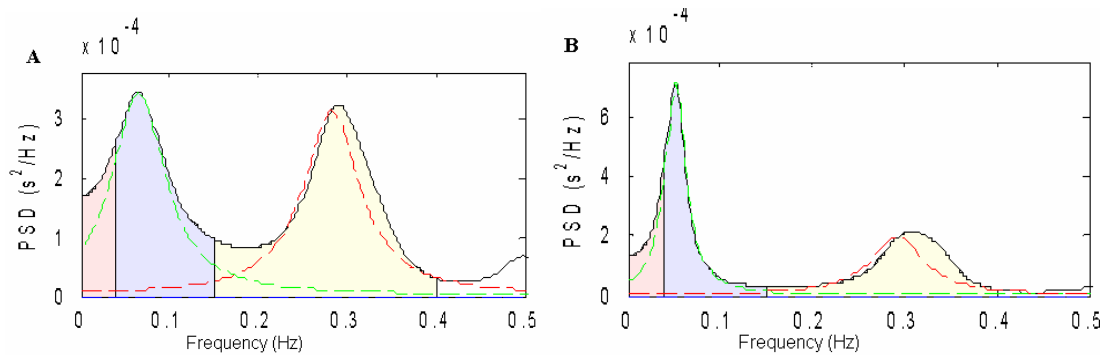

Fig 3 : Parametric spectrum (AR Model) of heart rate variability of a healthy subject calculated simultaneously from Pulse Wave Signal Peaks (A) and ECG R-wave Peaks (B). The Low Frequency (LF) and High Frequency (HF) peaks closely correspond in both figures.

The pulse wave signal was recorded from several regions on the body (e.g. Internal Carotid, Brachial, Radial artery). As observed in blood pressure waveform measured by other techniques there were distinct difference in characteristics of the shape of the pressure wave differential at different region. There exists a definite tie gap between two opposite wave while recording in carotid artery, which is absent when recorded from radial artery, and brachial artery signal has an intermediate shape.(Fig 4)

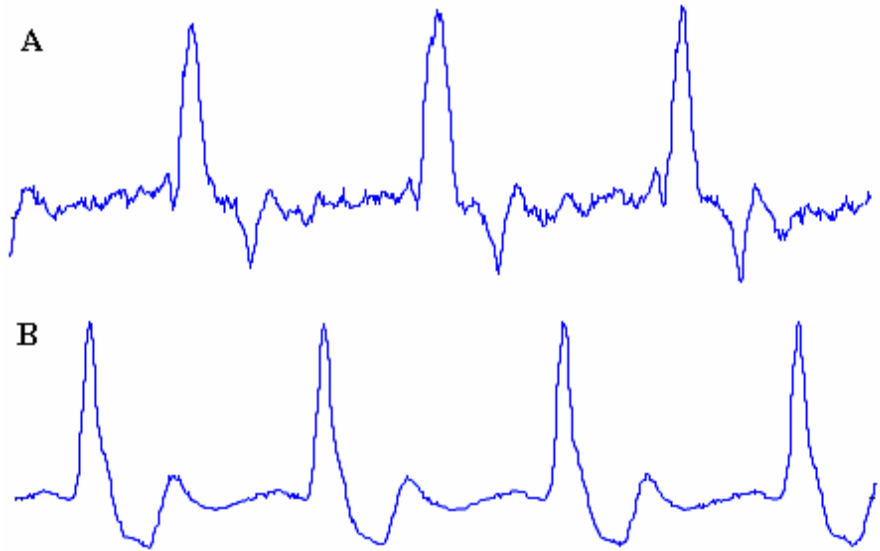

Figure 4 : Pulse wave signal recorded from the Carotid (A) and Radial (B) artery of same subject. The distinct plateau present between positive and negative peak in Carotid Artery recording was absent in Radial Artery.

Systemic blood pressure measurement was possible using the pulse wave signal obtained here. Pressure cuffs were applied in the arms of the subjects and signal from the radial artery of the same arm were recoded. Peak to peak amplitude from radial artery waveform was plotted against varying cuff pressure. The amplitude was seen to increase gradually with increase of cuff pressure from 0 mm of Hg to reach a peak when cuff pressure reaches the diastolic pressure of the subject, and to decrease in amplitude when cuff pressure is raised further. As the cuff pressure reaches the systolic blood pressure of the subject the pulse wave disappears. . Similar results were obtained when the cuff pressure is raised above the systolic BP ( depicted by disappearance of pulse wave signal) and slowly deflated and by applying a radial pressure cuff just distal to the pulse wave sensor.. (Fig5A). To obtain the pulse signal amplitude at different cuff pressure incremental step of 10 mm of Hg was used, restricted by the accuracy of barometric sphygmomanometer cuff (Fig 5B). The Blood Pressure estimated from the signal faithfully represented the Blood Pressure measured by auscultatory method (n=10).

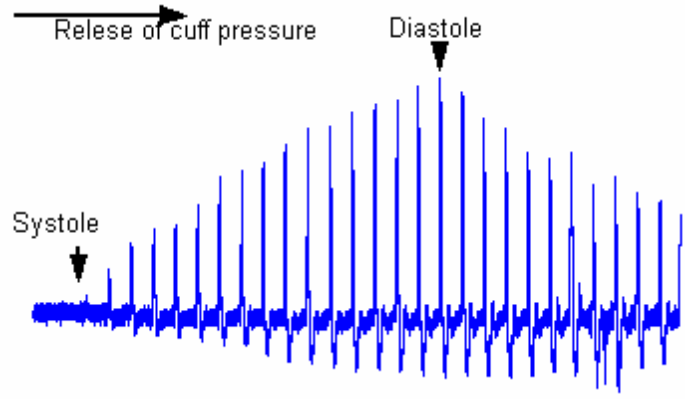

Figure 5 A: Shows the amplitude of pulse wave signal during slow deflation of radial cuff from a pressure well above systolic BP ( Characterized by absence of any pulse wave signal). The appearance of signal denotes the systolic pressure and the cuff pressure when the signal amplitude is maximum denotes diastolic pressure.

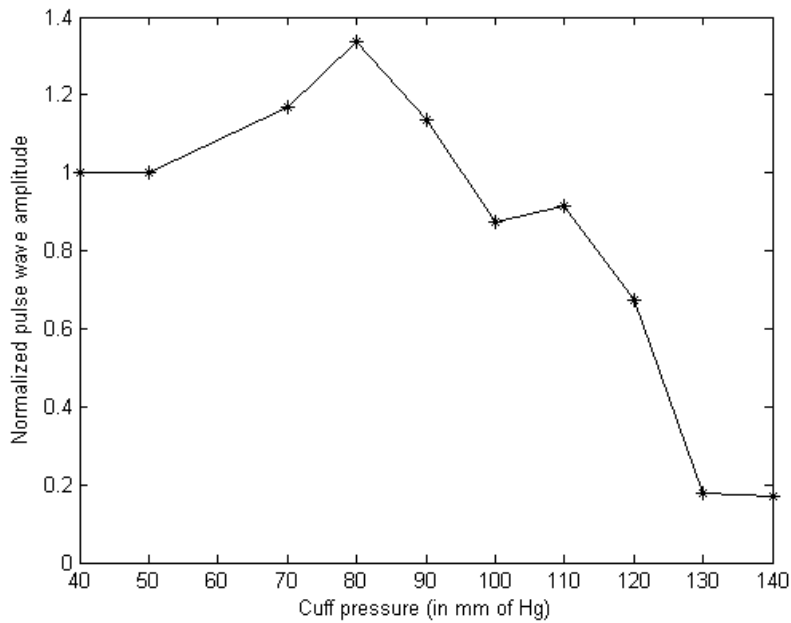

Figure 5B : Estimation of Blood pressure from pulse wave signal. The normalized amplitude of pulse wave signal of a healthy individual was plotted against arm cuff pressures at 10 mmHg increments. The pulse wave signal is absent from 130 mm Hg and above (where the plot is horizontal close to zero). Appearance of pressure wave occurred between 120 and 130 mm Hg, thus the calculated systolic BP is 125 mm Hg (avg of 120 and 130). Peak of amplitude is observed at 80 mm Hg (but it might have actually occurred anywhere between 70 and 90 mm Hg), hence calculated diastolic BP is 80 mm Hg. BP measured by auscultatory method was 124 / 74 mm Hg. (Here the accuracy of

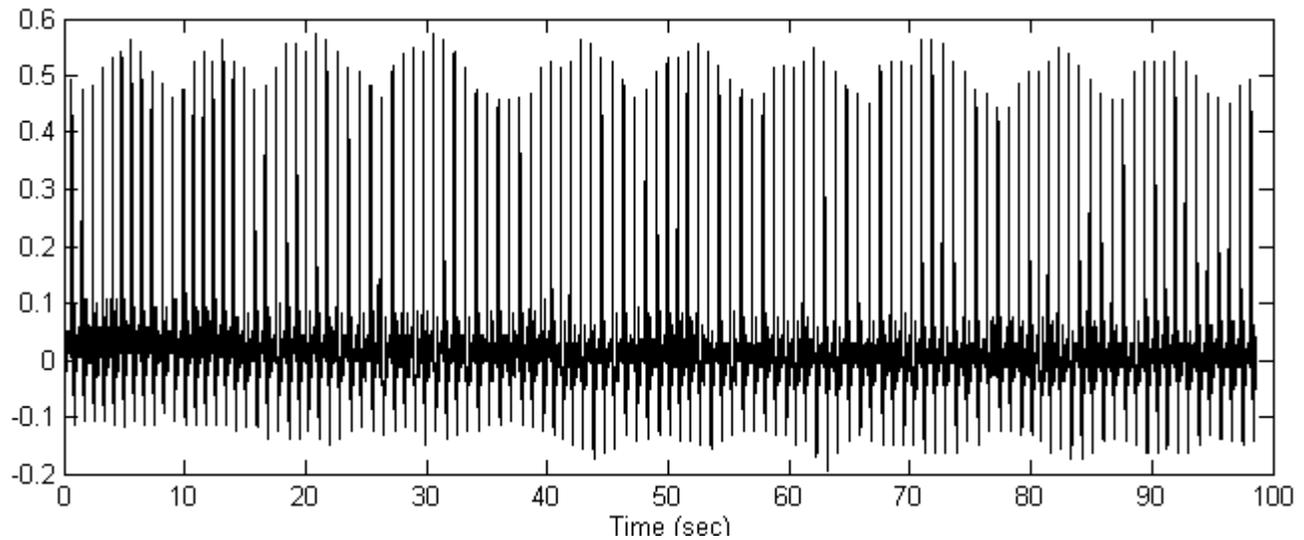

measurement could be substantially improved by continuous change in cuff pressure and simultaneous measurement of pulse wave amplitude, instead of using 10 mm Hg steps).

Pulse signal was recorded during different breathing maneuvers. Distinct modulation was observed when subject was asked to take deep breath. The positive peak of the pulse signal(which represents the rate of increase of blood pressure) reaches maximum amplitude at the height of inhalation and becomes minimum during peak of exhalation. The negative peak of the pulse wave also shows similar nature but less distinct.(Fig6)

Figure 6: Modulation of Pulse Wave Signal amplitude with deep respiration. The positive peak shows obvious changes in the amplitude. The negative peak also shows modulation but less obvious.

Signal recording was also done when the subjects were asked to perform Valsalva maneuver (Forced expiration with mouth and nostril closed). During the Valsalva the heart rate increased, the positive peak of the signal reduced and the negative peak is increased and there was appearance of a distinct second positive peak (which is very insignificant during normal breathing). Quick recovery of the peaks occurred after opening the glottis. Furthermore there was a reduction of the heart rate as expected from the baroreceptor reflex mechanism.(Fig-7)

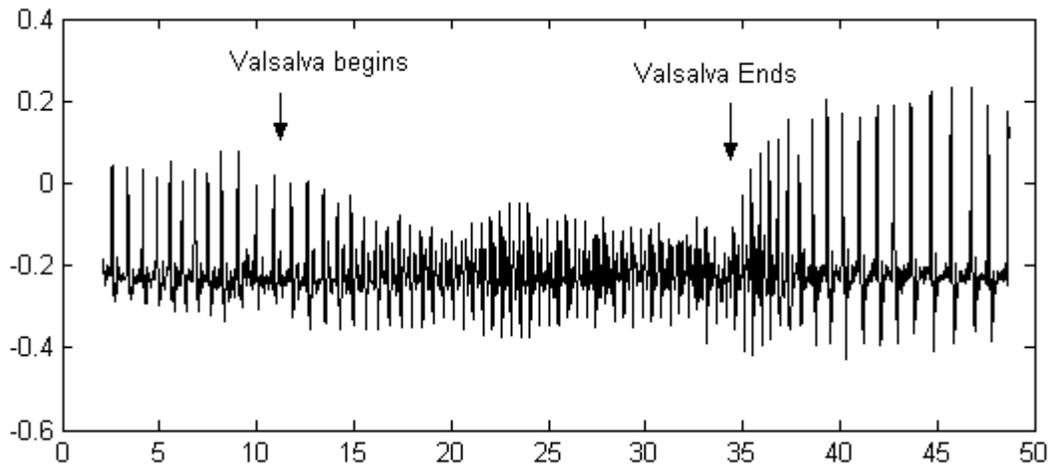

Figure 7 : Change in Pulse Wave Signal during Valsalva Maneuver. There was reduction in positive peak amplitude along with increased heart rate during the positive intrathoracic pressure. The amplitude and heart rate recovers as Valsalva ends.

A significant change in the pulse waveform was also observed when the subjects were asked to create forced negative thoracic pressure. Negative pressure was obtained by forceful effort to expand chest after a deep expiration and closing the glottis. Contrary to the valsalva maneuver finding in this case both the positive and the negative peak of the signal was severely diminished, almost to the noise level. A high reduction is also observed in the pressure wave amplitude obtained by passing the signal through the integrator circuit. Quick recovery occurs after release of the negative pressure. This proves a very significant reduction in the cardiac output occurs during the forceful negative thoracic pressure creation.(Fig 8)

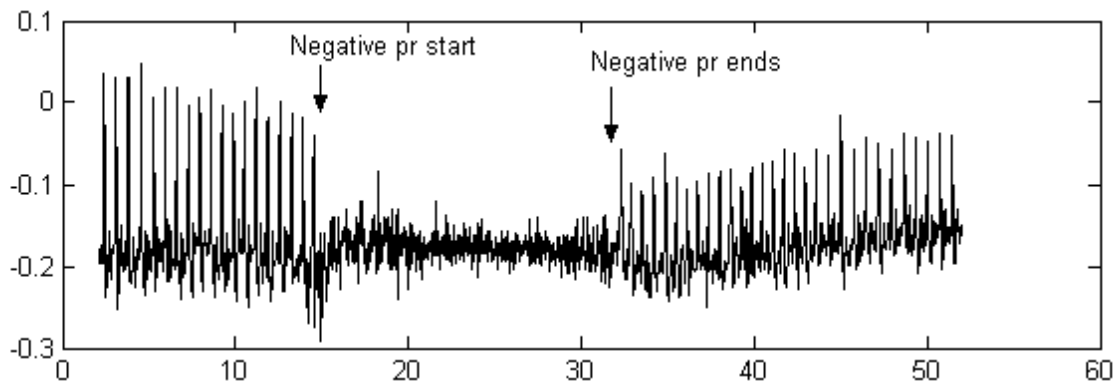

Figure 8: Change in Pulse Wave Signal amplitude during development of negative intrathoracic pressure. The signal amplitude almost disappears during height of negative pressure development.

**Discussions:**

The sensor described here shows a novel way of detection of arterial blood pressure waveform. The simplicity of recording procedure (Signal can be recorded by placing the sensor over a palpable artery and fixing it by a Velcro tape) might make it useful for wide application in Outpatient Department and Emergency.

The principal characteristic of this sensor is that it records the differentiated pressure wave signal which provides several advantages over other conventional procedures. (1) The blood pressure waveform can be retrieved form the pulse wave signal we obtain here by passing it through an integrator circuit, thus both the signals can be displayed concurrently. (2) The signals obtained here shows two distinct peaks (one positive and one negative), which represents the rate of change in the arterial diameter due to contraction of heart. The positive peak represents the increase in arterial pressure due to contraction of heart while the negative peak represents the sudden reduction in arterial pressure due to backflow of blood at the beginning of cardiac diastole. Thus if there are any change in the cardiac contractility due to acute pathology (for example in acute angina or silent myocardial infarction) without any change in the cardiac rate , rhythm and average blood pressure, there will be change in the shape and height of the peaks due to altered contraction pattern. Thus this measurement may help in diagnosis of such diseases where there is variability in contraction pattern of the heart without any other alteration, which can be difficult to diagnose otherwise by common methods such as ECG. (3) The signal is sensitive to diagnose the rhythm abnormality of the heart (as discussed in results section it is sensitive to detect irregular heart beat pattern in apparently fit individual). (4) Measurement of Blood pressure of an individual can be done from the signal. This will be particularly useful in development of noninvasive automated blood pressure wave monitoring system using the sensor. The simplicity of BP measurement procedure might draw one's attention. A cuff placed over the arm or wrist distal to the position of the pulse wave sensor is to be inflated up to the pressure when no signal comes. Then it is to be slowly deflated. The pressure at which first appearance of signal occurs denotes systole and the pressure at which it reaches maximum signifies diastole. (4) Using two sensors at the same time it will be possible to accurately determine and compare the pressure wave velocities at different parts of the body. Thus it may help diagnose peripheral obstruction of the flow or potential arteriosclerosis. (5) The ability of the sensor to detect the HPV with fair accuracy could enable one to get rid of ECG leads where simultaneous BP and HPV needs to monitored (e. g. in assessment of stress response or circadian "nondipping" in hypertensive patients).

In the present study a single sensor is placed over a pulsating artery but instead using an array of sensor might make it more sensitive and would not require the expertise to place the sensor properly over the artery. The same sensor can also be used to assess muscular contraction which is currently under study.

**Acknowledgement**: The authors are thankful to Dr.A.Mitra, and Mr. S.Ghosh for technical help.